# Observation of non-Abelian band topology without time-reversal symmetry

Yuze Hu[1,2,†], Mingyu Tong[1,7,†], Tian Jiang[2], Jian-hua Jiang[3,4,*], Hongsheng Chen[1,5,6,7,*], Yihao Yang[1,5,6,*]

[1] *Interdisciplinary Center for Quantum Information, State Key Laboratory of Extreme Photonics and Instrumentation, ZJU-Hangzhou Global Scientific and Technological Innovation Center, Zhejiang University, Hangzhou 310027, China.*

[2] *Institute for Quantum Science and Technology, College of Science, National University of Defense Technology, Changsha 410073, China*

[3] *Suzhou Institute for Advanced Research, University of Science and Technology of China, Suzhou 215123, China*

[4] *School of Physical Science and Technology & Collaborative Innovation Center of Suzhou Nano Science and Technology, Soochow University, Suzhou 215006, China*

[5] *International Joint Innovation Center, The Electromagnetics Academy at Zhejiang university, Zhejiang University, Haining 314400, China.*

[6] *Key Lab. of Advanced Micro/Nano Electronic Devices & Smart Systems of Zhejiang, Jinhua Institute of Zhejiang University, Zhejiang University, Jinhua 321099, China.*

[7] *Shaoxing Institute of Zhejiang University, Zhejiang University, Shaoxing 312000, China.*

[†] *These authors contributed equally to this work.*

[*] *yangyihao@zju.edu.cn; hansomchen@zju.edu.cn; jhjiang3@ustc.edu.cn*

**Abstract:**

Going beyond the conventional theory, non-Abelian band topology uncovers the global quantum geometry of Bloch bands with multiple gaps and thus unveil a new paradigm for topological physics. However, to date, all non-Abelian topological materials are restricted to systems with time-reversal symmetry ($\mathcal{T}$). Here, starting from a Kagome lattice inspired by Haldane model and designer gyromagnetic photonic crystals (PhCs), we show that $\mathcal{T}$ breaking can lead to rich non-Abelian topological physics, particularly the emergence of multigap antichiral edge states. Simply changing the magnetic flux of the Kagome lattice, or in-situ tuning the local magnetic field of the gyromagnetic PhCs, can lead to the unconventional creation, braiding, merging, and splitting of non-Abelian charged band nodes, alongside with the direct manipulation of the multigap antichiral edge states. Particularly, the quadratic point can be split into four Dirac points, a phenomenon unique in $\mathcal{T}$-broken systems. Our theoretical and experimental findings will inspire a new direction in the study of non-Abelian physics in $\mathcal{T}$-broken systems and open an unprecedent pathway for topological manipulation of electromagnetic waves.

**Main text:**

Topological theory unveils the physics underlying the exotic phenomena in insulators and semimetals, particularly through the connection between the bulk topology and the emergent boundary states. The latter, known as the bulk-boundary correspondence, is conventionally studied for a single topological band gap. In this context, the topological invariants are usually additive (i.e., Abelian topological invariants) [1-3]. In recent years, it is found that the physics of topological materials can be very rich and go beyond the conventional single band gap scheme, leading to challenges in understanding them [4-6]. Starting from understanding the intricate geometry and topology of nodal lines and nodal chains in three-dimensional (3D) crystals, non-Abelian topological band theory has uncovered the intriguing global structure of the Bloch wavefunctions of many bands with multiple band gaps, and thus unveils unconventional multigap topology, characterized by matrix-valued topological invariants, which goes beyond the previous theory of topological insulators and semimetals [7-19]. It turns out that such multigap topology is the key to fully understand the creation, annihilation, merging and splitting of the band nodes, revealing interesting effects beyond the Nielsen-Ninomiya theorem [20-22]. In this way, non-Abelian band topology fundamentally revolutionizes our understanding of topological phases. Yet, all previous studies have been limited to the systems with preserved $\mathcal{T}$, and the implications of non-Abelian band topology in the $\mathcal{T}$-broken systems remain uncharted.

Chiral edge states, initially discovered in a two-dimensional (2D) electron gas with a uniform magnetic flux in the 1980s, are one of the most striking phenomena in topological physics [23]. Lately, the well-celebrated Haldane model revealed that the chiral edge states can emerge intrinsically in a lattice with zero net magnetic flux, which signifies that $\mathcal{T}$ breaking plays a more fundamental role in forming chiral edge states [24]. Recently, it was shown that in a modified Haldane model, a pair of edge states propagating in the same direction at opposite boundaries, which are known as antichiral edge states [25-27]. These topologically protected antichiral edge states with suppressed backscattering are promising for energy-efficient electronics and light steering in photonic chips [27,28]. Up till now, the $\mathcal{T}$-breaking states, such as antichiral edge states, have been limited to the systems exhibiting scalar-valued Abelian topological invariants.

Here, we connect the two seemingly uncorrelated topics together, i.e., the antichiral edge states and the non-Abelian band topology. Inspired by Haldane model, we propose a Kagome lattice with staggered magnetic flux, in which multigap antichiral edge states originate from the non-Abelian topological charges. This Kagome lattice breaks both $\mathcal{T}$ and $C_{2z}$ rotation (i.e., the 180° rotation along the z direction) symmetry but preserves the combined symmetry $C_{2z}\mathcal{T}$. Thus, the non-Abelian band topology is still valid, while abundant topological phases emerge. Remarkably, the underlying physics can be demonstrated in a designer gyromagnetic photonic crystal (PhC). With near-field pump-probe measurements and in-situ tuning of the external magnetic field, we show that the multi-gap antichiral edge states can be manipulated by creating, moving, splitting (e.g., the splitting of a quadratic point into four Dirac points, unique in $\mathcal{T}$-broken systems), and merging non-Abelian charged topological band nodes in the Brillouin zone. These findings uncover the deep connection between antichiral edge states and non-Abelian topology and thus unveil an uncharted realm for topological phenomena and topological photonics.

Inspired by Haldane model, we start with a Kagome tight-binding model with tunable nearest-neighbor (NN) couplings whose amplitude is $t_1$ and phase is $\pm\varphi$ as illustrated in Figs. 1a-b. The three sites in each unit cell are labelled as 'A', 'B' and 'C'. In this model, the two corners of each unit cell have opposite magnetic fluxes due to the staggered phases in the NN couplings (see more details in Methods). This model has $C_{2z}\mathcal{T}$, while both $\mathcal{T}$ and $C_{2z}$ rotation symmetry are broken. The Hamiltonian in momentum space reads,

$$H(\boldsymbol{k}) = \begin{pmatrix} M_A & t_1 e^{-i\varphi} e^{-i\mathbf{k}\cdot\mathbf{a_1}} & t_1 e^{-i\varphi} e^{-i\mathbf{k}\cdot\mathbf{a_3}} \\ t_1 e^{i\varphi} e^{i\mathbf{k}\cdot\mathbf{a_1}} & M_B & t_1 e^{-i\varphi} e^{-i\mathbf{k}\cdot\mathbf{a_2}} \\ t_1 e^{i\varphi} e^{i\mathbf{k}\cdot\mathbf{a_3}} & t_1 e^{i\varphi} e^{i\mathbf{k}\cdot\mathbf{a_2}} & M_C \end{pmatrix} \quad (1)$$

where $\mathbf{k}=(k_x,k_y)$ is the momentum in the reciprocal space. $\mathbf{a_1}=(1/4,-\sqrt{3}/4)a_0$, $\mathbf{a_2}=(-1/2,0)a_0$, and $\mathbf{a_3}=(1/4,\sqrt{3}/4)a_0$ are the three real-space lattice vectors with $a_0$ being the edge length of a hexagonal unit cell. $M_j(j=A,B,C)$ represents a relative detuning (energy difference) associated with the three sublattice sites A, B and C, respectively. Here, we focus on the evolution of the non-Abelian topological semimetal states by varying the phase φ, while keeping the hopping amplitude as $t_1=1$. We find that, as shown in Fig. 1c (corresponding energy bands are shown in Figs. S1-S2 of supplementary text), the creation, splitting and merging of the non-Abelian band nodes are extremely rich and follows rules that are different from time-reversal symmetric systems studied previously [6,10,11]. We remark that the model in Eq. (1) can be realized in a gyromagnetic PhC where the hopping phase can be in-situ controlled by the external magnetic field (Figs. 1d-e), which will be specified in detail below.

Compared with the well-celebrated Haldane model composed of two sublattices [24,27], our model is a Kagome lattice composed of three sublattices and features staggered magnetic flux. As $C_{2z}\mathcal{T}$ is preserved in this model, the Bloch Hamiltonian can become real-valued in an appropriate basis. Most importantly, the Bloch eigenstates at each wavevector define a 3D orthonormal frame (up to a sign ambiguity). The evolution of the frame in the reciprocal space around a band node yields the non-Abelian frame charge as the topological invariant of such a node. These non-Abelian frame charges belong to the quaternion group's conjugacy classes $\mathbb{Q} \in \{1, \pm i, \pm j, \pm k, -1\}$ [9,29]. The trivial charge corresponds to $\mathbb{Q}=1$. The quaternion group's fundamental multiplication rules give $i \times j = -j \times i = k$, $j \times k = -k \times j = i$, $k \times i = -i \times k = j$, as well as the squares of the imaginary units $i^2 = j^2 = k^2 = -1$ [4,9,10]. These rules dictate the "reaction laws" when the band nodes merge as they give the total quaternion charge of a set of band nodes. For instance, a set of band nodes can be annihilated (i.e., gapped out) only when their total quaternion charge is $\mathbb{Q}=1$.

The evolution of the non-Abelian frame charges of the band nodes in the parameter space $(k_x, k_y, \varphi)$ in Fig. 1c gives the phase diagram of the non-Abelian topological semimetal and illustrates elegantly the topological transitions due to the splitting and fusion of the non-Abelian band nodes. Unlike the previous studies with varied hopping amplitude [10,11,22], such a process is triggered by the tuning of the hopping phase accompanied by $\mathcal{T}$ breaking and the multiband structure is tuned substantially (see Figs. 1f-i).

We start the tuning process with the familiar case where no magnetic flux is applied, that is, $\varphi=0$. In this case, there are a quadratic node at the Γ point in the first bandgap (denoted henceforth as 'gap I') and linear (Dirac) nodes at the K and K′ points in the second bandgap (denoted as 'gap II'). The Dirac nodes in gap I have frame charges $\mathbb{Q}=-1$, whereas the quadratic node in gap II has frame charge $\mathbb{Q}=\pm i$ (Fig. 1f). By continuously decreasing the strength of the NN coupling and increasing the strength of the magnetic flux, rich non-Abelian topological phases can emerge. First, we find that the quadratic node is split into four Dirac nodes with one frame charge $\mathbb{Q}=-k$ at the Γ point and three frame charges $\mathbb{Q}=+k$ out of the Γ point. These Dirac nodes can be created together because they have a total frame charge of $\mathbb{Q}=-1=(+k)^3(-k)$, demonstrating a case where Dirac nodes can be created or annihilated together according to quaternion operations.

Constrained by the three-fold rotation crystalline symmetry, three Dirac nodes with frame charges $\mathbb{Q}=+k$ are located on three ΓK' lines in the 2D Brillouin zone (Fig. 1g) and moving towards the K' point with further tuning magnetic flux. Merging three Dirac nodes in gap II and merge together with the $+i$ frame charge Dirac node in gap I at the K' point leads to a linear triply-degenerate point with frame charge $\mathbb{Q}=-j$ (Fig. 1h) [30]. This process can be regarded as the merging of one Dirac node in gap II and three nodes in gap I, leading to a total frame charge of $\mathbb{Q}=(+i)(-k)^3=-j$ for the triply-degenerate point at the K' point. By further tuning the flux, the triply-degenerate point is split into four Dirac nodes with one frame charge $\mathbb{Q}=+k$ at the K' point and three frame charges $\mathbb{Q}=(+i)^3$ along the K'K lines according to the quaternion operation $\mathbb{Q}=-j=(+i)^3(+k)$ (Fig. 1i).

Further tuning the system merges three frame charges $\mathbb{Q}=(+i)^3$ with one frame charge $\mathbb{Q}=-i$, leading to a quadratic node with $\mathbb{Q}=(+i)^3(-i)=-1$ at K point. Meanwhile, such quadratic node would split into three $+i$ charged Dirac nodes moving toward Γ point and one $-i$ charged node fixing on K point. Then, an operation of $\mathbb{Q}=-j=(-k)(+i)^3$ is realized at Γ point when $\varphi=\pi/2$. A similar evolution process is realized when $\varphi$ increases from $\pi/2$ to $\pi$, whose details are illustrated in Fig. 1c and Fig. S2 in Supplementary Text. It is noticeable the hopping amplitude changes from 1 to −1 when $\varphi$ varies from 0 to $\pi$. In the view of the band structure, a quadratic point transfers from Gap I to Gap II by simply increasing the $\mathcal{T}$ breaking strength in our proposed model. During the whole tuning process, the coalescence of the Dirac nodes in Gap-I and Gap-II leads to the frame charge $\mathbb{Q}=-1$ emerging at either Γ, K or K' points.

At this stage, it is important to summarize the salient features in the above. First, by breaking the $\mathcal{T}$, the quadratic band touching at the Γ point is no longer protected. This leads to node splitting and elegant evolution which constructs multiband nodal links in the three-dimensional synthetic space. Second, by simply varying the magnetic flux in Kagome lattices, the evolution process includes the creation, annihilation, merging and splitting of nodes in two gaps, resulting in several quaternion operation of non-Abelian charges. Even though the emergent triply-degenerate point leads to the transfer of the non-Abelian charges between different bandgaps, the splitting of a quadratic point plays a central role in creating additional tunable non-Abelian charges. Third, we find that, in our model, the quadratic point in Gaps I and II can be interconverted by introducing phase difference in the hopping coefficients through $\mathcal{T}$ breaking. Such features, which cannot be understood via conventional $\mathcal{T}$-preserving multiband non-Abelian models where the quadratic point is robustly protected [11], faithfully reveal rich phase diagrams of the $\mathcal{T}$-broken multigap topology.

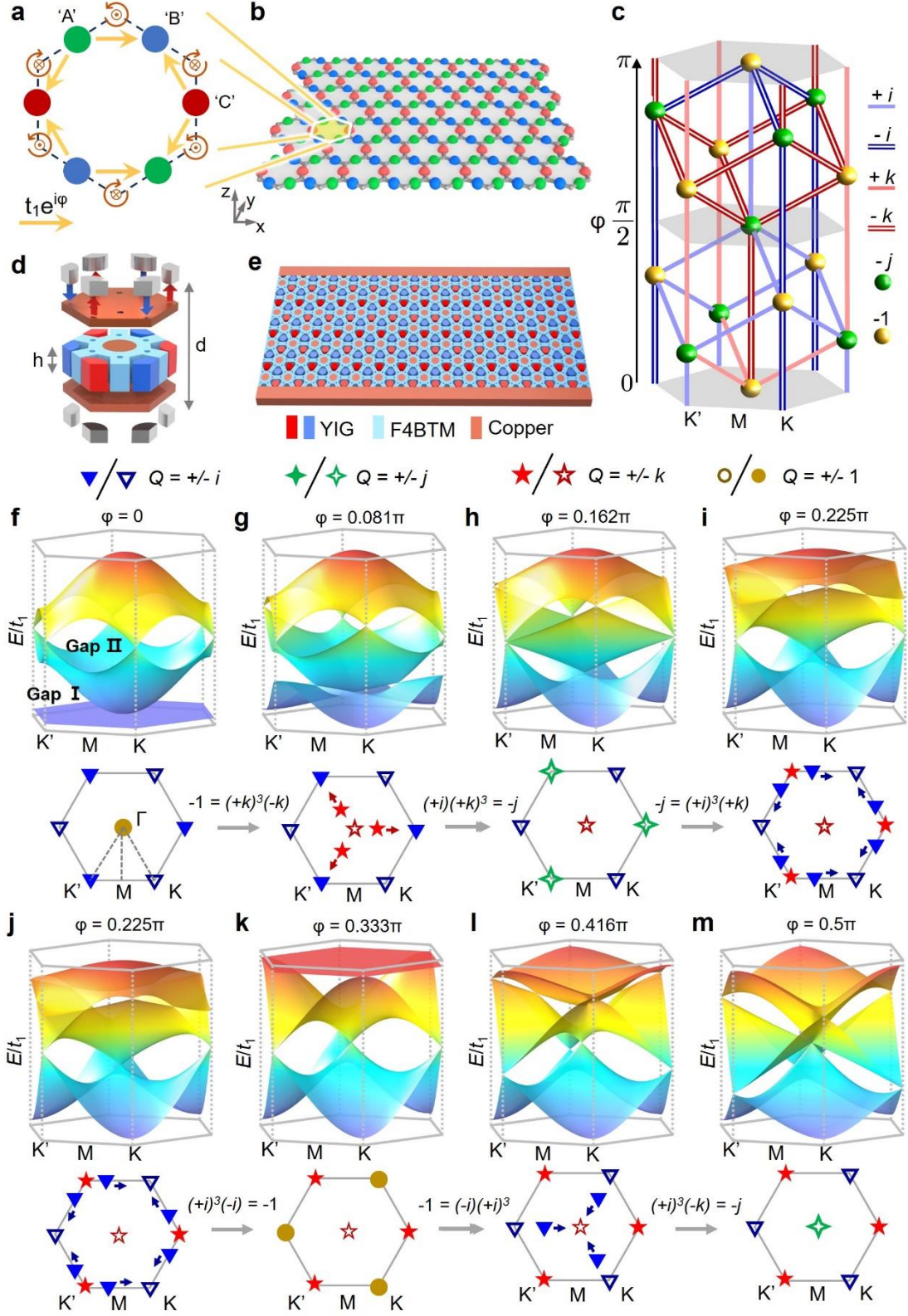

**Fig. 1. Non-Abelian topological charges in a Kagome lattice with staggered magnetic flux. a**, Schematic view of the Kagome tight-binding configuration with staggered magnetic flux, where the nearest neighbor hopping is considered. **b**, Tight binding lattice consisting of three atoms in each unit cell. **c**, Phase diagram depicting

the evolution of non-Abelian topological charges as a function of the magnetic flux. A unit cell (**d**) and honeycomb arrangement (**e**) of the gyromagnetic PhC consisting of resonators made of F4BTM that are interleave with YIG cylinders. Varied magnetic flux can be realized by changing the distance $d$ between upper and bottom magnets. $h=5\,\text{mm}$ is the thickness of PhCs. **f-m**, Evolution of the band structures (upper panels) and the nodes in gaps I and II (bottom panels) by continuously increasing the strength of magnetic flux from $\varphi=0$ to $\varphi=0.5\pi$ in the diagram of (**c**). Solid (open) blue triangles label the Dirac nodes in gap I with quaternion frame charge $\mathbb{Q}=i(-i)$. Solid (open) red quadrangular stars label the Dirac nodes in gap II with frame charge $\mathbb{Q}=j(-j)$. Solid (open) green pentagrams denote the linear triply-degenerate points with frame charges $\mathbb{Q}=k(-k)$. Solid (open) brown circles label the quadratic nodes with frame charge $\mathbb{Q}=-1(+1)$.

The evolution of bulk topological charge configurations reveals theoretically predictable variations in edge states across tuning parameters, offering an insight into the nature of topological band nodes. These nodes significantly affect edge states behavior in Gap I and Gap II, where a momentum-dependent, quantized Zak phase appears. Given the periodicity of Brillouin zone, the quantized Zak phase of each band can be calculated by considering its subsystems formed by 1D loops of constant $k_x$ with a parameter $k_y$ scanning one period [31]. This phase in the corresponding gaps undergoes a transition between 0 and $\pi$ upon encountering an odd number of Dirac nodes; however, it remains unchanged when crossing a quadratic node or a pair of Dirac nodes with same frame charges [32]. A triply-degenerate point characterized by a frame charge $\mathbb{Q}=\pm j$ effectively represents an odd number of Dirac nodes in both energy gaps. Under the conditions in our model with Zak gauge, a Zak phase of 0 ($\pi$) results in the appearance of time-reversal breaking induced antichiral edge states in Gap I (II) (see details in supplementary text). Interestingly, for a given edge boundary, the Zak phase here is quantized by the combination of the mirror symmetry and $\mathcal{T}$, although none of them is preserved [33]. Such symmetry becomes concrete when considering the ribbon structure containing the edge boundary, as shown in Fig. 2. The creation, braiding, merging, splitting and annihilation of the non-Abelian band nodes determines the Zak phase in each partial band gap, which together with the energy shift of these band nodes determine the evolution of the edge states. In our case, the edge states can be engineered to have opposite group velocities in different band gaps, leading to versatile topological steering of light.

We consider a zigzag nanoribbon of the Kagome lattice infinitely long along $x$-direction but having a finite width along $y$-direction as shown in Fig. 2a. Here, the onsite energy of three atoms in the bulk is set as $M_A = M_B = M_C = 5t_1$, while the onsite energy at zigzag edges is $6t_1$ to ensure the edges states locate in the middle of the bulk bandgap. The projected tight-binding band structures of this ribbon are plotted for different values of magnetic flux ($\varphi$) in Fig. 2b, f, and j. The red and blue curves represent the edge states inside Gap I and Gap II, respectively. Since our non-Abelian model is subject to the oppositely imposed staggered magnetic fluxes, the Dirac nodes shift frequencies, causing the tilted dispersion of two edge states. The antichiral edge states are topologically protected and spatially separated from the bulk states. These states are different from the edge states in a conventional modified Haldane model, because the emergences and disappearances of two edge states are highly dependent on the multigap non-Abelian frame charges. Thus, their characteristics can be tuned by external magnetic flux (schematically shown in Figs. 2 a, e, and i).

For $\varphi = 0.081\pi$, there are two types of antichiral modes confined at a given energy in the bulk Gaps I and II, respectively, as shown in Fig. 2b. In Fig. 2c, it is visible that odd numbers of Dirac nodes in a single gap act as the termination of edge states by switching the Zak phase between 0 and $\pi$. In this case, the edge states with $E = 5.65t_1$ and $E = 3.75t_1$ are antichiral. The corresponding full time-dynamic calculations in Fig. 2d serve as a testament to the validity of above-mentioned statements. Increasing the magnetic flux when $\varphi = 0.162\pi$, the triply-degenerate points with frame charge $\mathbb{Q} = -j$ play the same role as the Dirac nodes but act as the termination of edge states of both gaps (Fig. 2g). Here, both antichiral edges states stand isolated in the bulk gaps and coexist with opposite group velocities, as shown in Figs. 2f and h. Further tuning the magnetic flux with $\varphi = 0.225\pi$, the Dirac nodes with $+i$ frame charge moves from K' to K point, leading to the shortening of upper edge states in Gap II. Meanwhile, the nodes ($\pm k$ frame charges) in Gap I maintain positional immobility. As a result, the upper antichiral edge states gradually disappear by leaving the bottom edges along, as illustrated in Figs. 2j, and m.

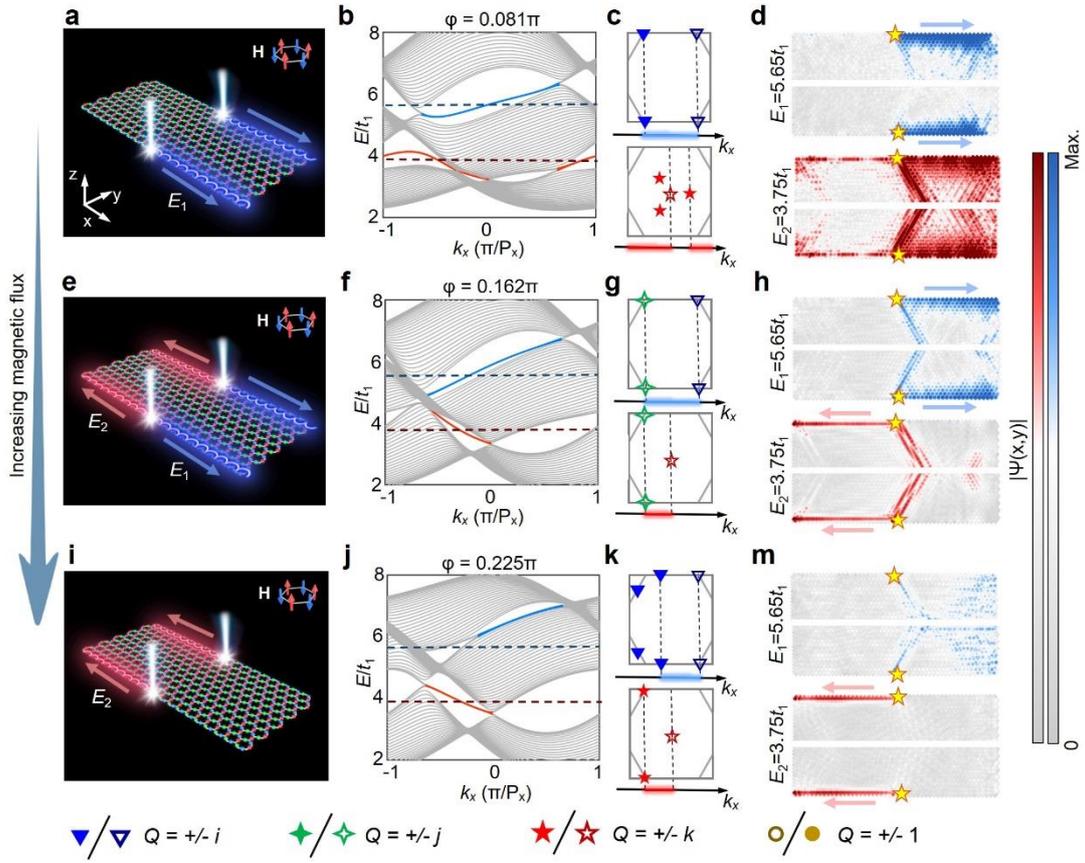

**Fig. 2. Multigap antichiral edge states from the non-Abelian topological charges**. **a, e, i**, Schematic representations of the variation of multigap antichiral edges as a result of bulk charge phase transitions. **b, f, j**, Projected energy dispersions for structures that are finite in the *x*-direction but periodic in the *y*-direction. Gray curves represent bulk states and blue (red) ones represent edge states in gap II (I). **c, g, k**, Upper (bottom) panel is the topological node arrangement within bandgap II (I) throughout the entire Brillouin zone. Mapping of Dirac nodes within bandgap I (II) is identified using red (blue) symbols. Red (blue) segments aligned along the $k_x$-axis mark the regions that possess a 0 ($\pi$) Zak phase within the Gap I (II). **d, h, m**, Time-dynamic calculations for a ribbon used with an excited point at the edge center. The excitation point oscillates with the energy of $5.65t_1$ (upper panels) and $3.75t_1$ (bottom panels), corresponding to the energies of antichiral edge states in gap II and I, respectively.

Based on above, we design a PhC to bridge the gap between theoretical expectations and experimental realizations. Recently, gyromagnetic PhCs have been demonstrated as a versatile platform for the study of antichiral topological phenomena [1,26,34]. The model necessitates control of magnetization via gyromagnetic cylinders at the corners of a honeycomb lattice and another three resonators to represent 'A', 'B', and 'C' atomic species. We adopt the configuration as schematically illustrated in Figs. 1d-e. The red and blue cylinders are yttrium iron garnet (YIG) ferrite materials biased in opposite magnetic direction, similar to previous demonstrations inspired by Haldane model. Importantly, in the current configuration, the nanoceramic-filled laminates reinforced with woven fiber-glass (F4TBM) with permittivity 3.5 is inserted into the crystal to form Kagome lattice and one copper cylinder at each unit-cell center is used to block the long-range hopping. Schematics and dimensions of the PhC are shown in Fig. S4 (see details in supplementary text). In particular, such a configuration gives access to measure the spectra of edge states with a flexible choosing of biased magnetic field strength without changing lattice structure, which is advantageous for the in-situ control of multigap non-Abelian topology according to bulk-edge correspondence.

Figure 3a shows the simulated bulk band structure of the Kagome-like gyromagnetic PhC without magnetic flux, which hosts two Dirac points and one quadratic point (marked by blue triangles and brown spheres). Due to the $C_{2z}\mathcal{T}$ of the lattice structure, both eigenstates of the tight-binding model and the PhC ($E_z$-polarized field distribution) at Γ point are real values. To render the concept more tangible, we invoke a comparative illustration that analogizes the tight-binding model by dividing the PhC region within one-unit cell into three distinct domains (marked as 'A', 'B', 'C' shown in left panel in Fig. 3e). The eigenmodes of three bands at the Γ point shown in Fig. 3a are visible in Fig. 3e. By performing the integrations across the three distinct spatial domains, the eigenstates are described by $3\times1$ matrix similar to a tight-binding eigenstate. Then, the eigenstate comparison between PhC and tight-binding is shown in Fig. 3f, reflecting a strong consistency. Such feature unambiguously validates the suitability of the designed PhC for mimicking Kagome lattice structures, establishing its capability to faithfully reproduce key non-Abelian phenomena.

We now proceed to demonstrate the evolution of non-Abelian frame charges in PhCs by tuning the oppositely staggered external magnetic flux strength. It is found that the bands in both gaps start to tilt when magnetic field is applied. When $\mu_0 H_z = 20$ mT, the quadradic point with frame charge $\mathbb{Q}=-1$ at Γ point is split into one Dirac node with $\mathbb{Q}=-k$ at Γ point and three nodes with $\mathbb{Q}=+k$ along ΓK' directions (Fig. 3b).

As the strength is incremented further to 30 mT, a triply-degenerate point with $\mathbb{Q} = -k$ appears at K' point by merging three $+k$ Dirac nodes with one $+i$ node (Fig. 3c). Then, the triply-degenerate point is converted into a $+k$ Dirac node and three $+i$ nodes moving to K point (Fig. 3d). The corresponding eigenstate frame spheres for each Dirac nodes are plotted in Fig. 3g, which give solid evidence of the non-Abelian charge for each Dirac nodes. The simulation results depicted in Fig. 3 directly instantiate the theoretical prediction of Dirac nodes evolution outlined in Fig. 1, thereby laying a foundation for experimental realizations of anticipated topological characteristics in the designed system.

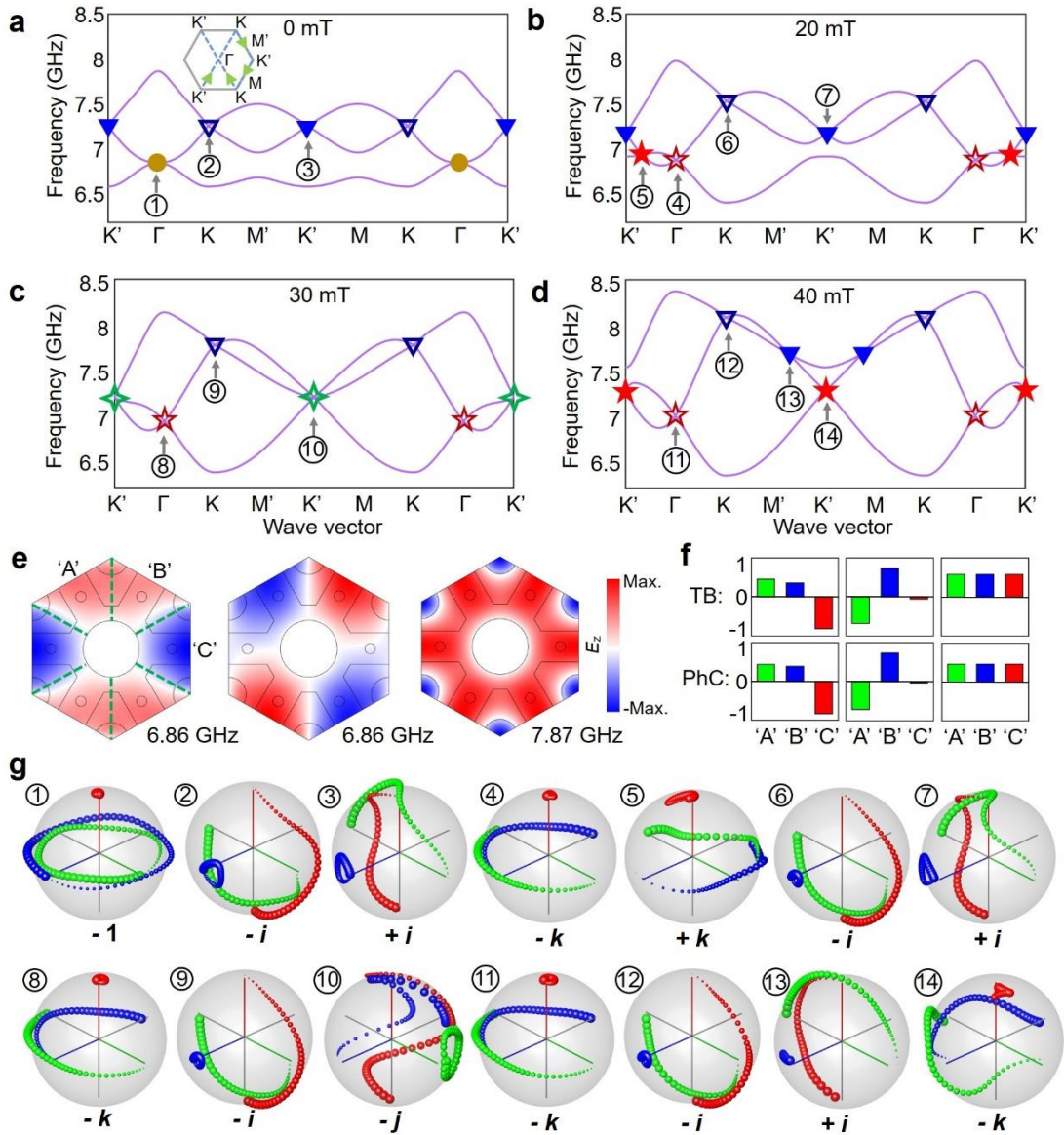

**Fig. 3. Design of non-Abelian gyromagnetic PhCs. a-d,** Band structures at different moments plotted along the high symmetry lines of the hexagon BZ with various

magnetic strengths applied on the YIG cylinders. The corresponding non-Abelian topological charges are highlighted on the Dirac nodes. **e**, Eigenfield distributions of three PhC states at the Γ point in (**a**). **f**, Eigenstate amplitudes of atoms 'A', 'B', and 'C' calculated from tight-binding model (upper panel) and PhC eigenmodes (bottom panel). **g**, Simulated eigenstate frame spheres for each node in (**a-d**), marked from ① to ⑭. The first, second and third bands are coloured as red, green and blue, respectively. The direction of increasing line width indicates circling the node clockwise.

We validate the theoretical framework of the nonconventional non-Abelian charge evolution by performing experiments on a ribbon-shaped PhC as shown in Fig. 4a. Copper cladding has been introduced near the upper and lower edges as perfect electric conductor (PEC) boundaries to prevent leakage of electromagnetic waves into the surrounding environment. The *x*-direction edges are covered with microwave absorbers. Two plates of permanent magnets with oppositely-biased staggered field are placed above and below the sample, whose distance can be flexibly controlled to vary the magnetic flux strength. The measured magnetic flux strength ($\mu_0 H_z$) at the middle plane between the two magnet plates is shown in Fig. 4b, which is highly dependent on the distance. When a source is located at the middle along *x*-direction and near two edges in *y*-direction, the excited edge states would propagate along ±*x*-directions. Via the Fourier transformation of the detected wavefunctions at each excitation frequency, we thus obtain the dispersions of the electromagnetic Bloch waves. Figs. 4c-k show that the measured electromagnetic band structures agree excellently with the theoretical predictions (details are provided in the Fig. 2), especially when the variation of magnetic flux strength is considered. The observed results in the PhC provide the first demonstration for the associated conversion and transfer of antichiral edge states in different bandgaps, which are simply tuned by the external magnetic flux. From material science aspects, all band structures in such Haldane-like Kagome lattice are unconventional dispersions. Thus, the discovery of these unique band structures in Kagome materials underscores their multifaceted physical attributes, which not only open new avenues for fundamental scientific inquiry—such as the study of emergent strongly correlated quantum phenomena within Kagome lattice frameworks—but also present promising prospects for technological applications, exemplified by the potential to engineer Dirac materials with tunable band characteristics enabling sophisticated wave control mechanisms.

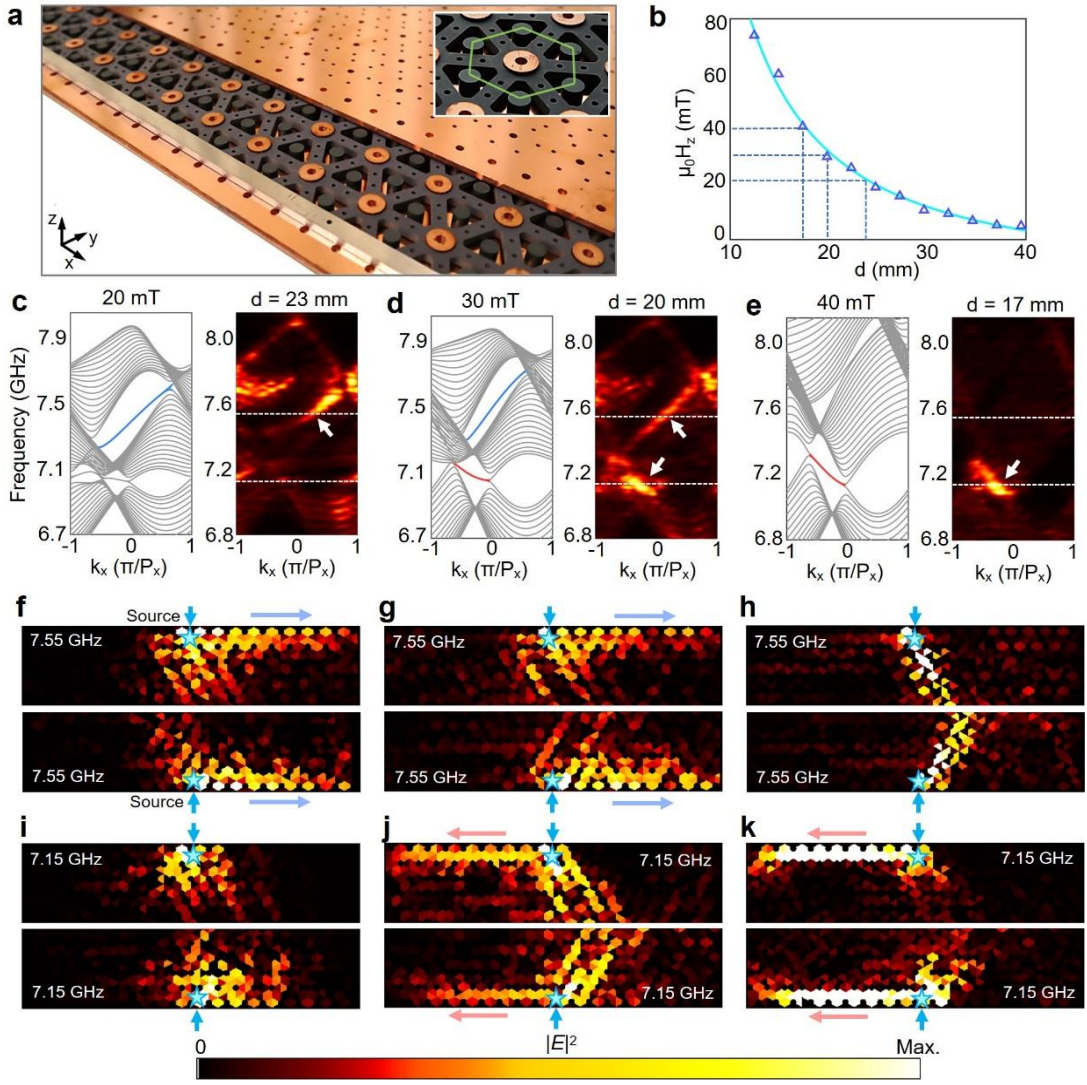

**Fig. 4**. **Observation of multigap antichiral edge states in a gyromagnetic PhC with in-situ controllable magnetic fields. a**, Experimental implementation of the non-Abelian gyromagnetic PhCs. The PhCs are placed in a parallel-plate copper waveguide with distance of 5 mm. The thickness of the upper and bottom waveguide plates is 1.5 mm. Two panels of permanent magnets of diameter 8 mm and height 3 mm are placed above and below the copper plates with varied distances to in-situ tune the magnetic field, as displayed in Fig. 1d. The upper copper layer is glided two lattices away from their edges for photographing and a unit cell is framed by green lines in the inset. **b**, Magnetic strength as a function of the distance between two permanent magnets. **c-e**, Projected energy bands as a function of magnetic strength. Left panels: numerically calculated band structures for a ribbon with 22-unit cells in the $y$-direction. The red and blue lines belong to antichiral edge states. Right panels: experimentally measured edge-

state band structures from Fourier transform. Measured electric field distributions for edge states (**f-h**) in gap II at the frequency of 7.55 GHz and (**i-k**) in gap I at the frequency of 7.15 GHz.

**Conclusions and outlook**

We have thus provided a theoretical and experimental framework that connects the antichiral edge states with the non-Abelian band topology in a Kagome lattice with broken $\mathcal{T}$ and $C_{2z}$ but preserved $C_{2z}\mathcal{T}$. By tuning the magnetic flux, we uncover non-Abelian topological phase transition, including the creation, annihilation, merging and splitting of non-Abelian charged band nodes. Particularly, the topologically stable quadratic point in $\mathcal{T}$-preserved systems can be split into four Dirac points, revealing the unconventional non-Abelian topological physics in $\mathcal{T}$-broken systems. As the multigap antichiral edge states originate from the non-Abelian topological charges, the magnetic-flux-induced phase transition provides a means to control the multigap antichiral edge states that serve as excellent transport channels. Most importantly, this Haldane-model-inspired Kagome lattice can be readily implemented using gyromagnetic PhCs, whose staggered external magnetic fields can be in-situ controlled. It would be interesting to implement our model in other physical systems, such acoustics [11,35,36], cold atom [37,38], condensed matter [2,5,39] systems, and study higher-order non-Abelian groups [15]. Moving forward, our findings not only apply the non-Abelian band topology to $\mathcal{T}$-broken systems, but also expand the Haldane model characterized by Abelian charges into multiband model characterized by non-Abelian charges, unveiling an uncharted realm for topological phenomena and topological photonics.

**Methods**

The Methods section includes details used for the simulations and experiments.

**Simulations**

In this work, numerical results of PhCs are obtained using the RF module of COMSOL Multiphysics. The bulk band structure simulation employs a hexagonal unit cell under periodic boundary conditions in all dimensions. To calculate edge dispersions, we employ a 1×22 supercell, imposing periodic boundaries along the *x*-axes and Perfect Electric Conductor (PEC) conditions along *y*-direction. Copper plates and permanent magnets are modeled as PECs. Gyromagnetic materials, fully magnetized, exhibit a permeability tensor $[\mu_r] = \begin{bmatrix} \mu_m & \pm j\mu_k & 0 \\ \mp j\mu_k & \mu_m & 0 \\ 0 & 0 & 1 \end{bmatrix}$, where

$\mu_m = 1 + \omega_m(\omega_0 + i\alpha\omega)/[(\omega_0 + i\alpha\omega)^2 - \omega^2]$, $\mu_k = \omega_m\omega/[(\omega_0 + i\alpha\omega)^2 - \omega^2]$,

$\omega_m = \mu_0\gamma M_s$, $\omega_0 = \gamma\mu_0 H_z$. The corresponding parameters are highly dependent on the external magnetic field $H_z$ along $z$-axis. Here, gyromagnetic ratio is $\gamma = 1.759 \times 10^{11}$ C/kg, damping coefficient is $\alpha = 0.0088$, and operational frequency is $\omega$. A weak material dispersion within the operational frequency range due to the operation frequencies being distant from the resonant frequency of the gyromagnetic material. We then detail the calculation of eigenstate trajectories revealing magnetic flux-dependent nodal quaternion-valued frame charges (Fig. 3). A 3D synthetic space $(k_x, k_y, \mu_0 H_z)$ is constructed by combining the 2D Brillouin zone and magnetic flux parameter. For each node, we trace an anticlockwise circle of radius 0.15, fixing a common basepoint $\mathbf{P}_0(k_{x0}, k_{y0}, 0)$ for consistency. The path follows: moving from $\mathbf{P}_0$ close to the node, incrementing magnetic flux to identify start point $\mathbf{A}_0$, then circling node anticlockwise. Trajectories of eigenstates from the first, second, and third bands projected onto $\mathbf{A}_0$ are depicted by blue, green, and red curves, respectively.

**Experiments**

In the experimental setup, commercial yttrium iron garnet (YIG) ferrites and Sm$_2$Co$_{17}$ permanent magnets are utilized to break $\mathcal{T}$. YIG ferrite cylinders have radius of 3.3 mm and heights of 5 mm, with a saturation magnetization M$_s$ = 3000 Gauss. Permanent magnets (3-mm-thick and 4-mm-radius) generate an approximately 0.3 Tesla uniform external magnetic field to magnetize the YIG rods. The permanent magnets are affixed to two parallel plates, positioned in precise alignment with the YIG ferrite rods. Manual manipulation is employed to regulate the separation distance between pairs of oppositely staggered magnetic plates with precision increments. Thus, the resultant magnetic field strength within the YIG ferrite rods is controllably modulated. In experimental measurements, a vector network analyzer (VNA) is interfaced with dual microwave dipole antennas serving as the source and detector, respectively. The local electric field's amplitude and phase information across the entire frequency range of interest are then acquired using the VNA. For the edge states dispersion measurement, the source is horizontally inserted into the upper copper plane through the air holes of the 2D PhC (six predefined holes per unit-cell). By scanning the sample in a point-wise manner, complex electric field distributions across the sample's surfaces and interior are mapped. Then, projected bulk and edge band structures are derived via 2D Fourier transformations applied to the measured complex electric field data at each frequency.

**Acknowledgements**

The work was sponsored by the Key Research and Development Program of the Ministry of Science and Technology under Grants 2022YFA1405200 (Y.Y.), No. 2022YFA1404704 (H.C.), 2022YFA1404902 (H.C.), and 2022YFA1404900 (Y.Y.), the National Natural Science Foundation of China (NNSFC) under Grants No. 62175215 (Y.Y.), No. 61975176 (H.C.), No. 62305384 (Y.H.), and No. 62305298 (M.T.), the China National Postdoctoral Program for Innovative Talents under Grants No. BX20230310 (M.T.), the Youth Innovation Talent Incubation Foundation of National University of Defense Technology under Grants No. 2023-lxy-fhij-007 (Y.H.), the Key Research and Development Program of Zhejiang Province under Grant No.2022C01036 (H.C.), the Fundamental Research Funds for the Central Universities (2021FZZX001-19) (Y.Y.), and the Excellent Young Scientists Fund Program (Overseas) of China (Y.Y.).

**Author contributions**

Y.H., Y.Y. and M.T. created the design. Y.H., M.T. and Y.Y. designed the experiment. Y.H. and M.T. fabricated samples. Y.H. carried out the measurement with the assistance from M.T. and Y.H., M.T. analyzed the data. M.T. performed simulations, J-H.J., Y.Y., T.J., and H.C. provided the theoretical explanations. Y.H., M.T. wrote the manuscript with the input from Y.Y., J-H.J., and Y.Y., T.J., H.C. and J-H.J. supervised the project. All authors contributed extensively to this work.

**Competing interests**

The authors declare no competing interests.

# Supplementary Materials for

# Observation of non-Abelian band topology without time-reversal symmetry

The PDF file includes:

Figs. S1 to S4

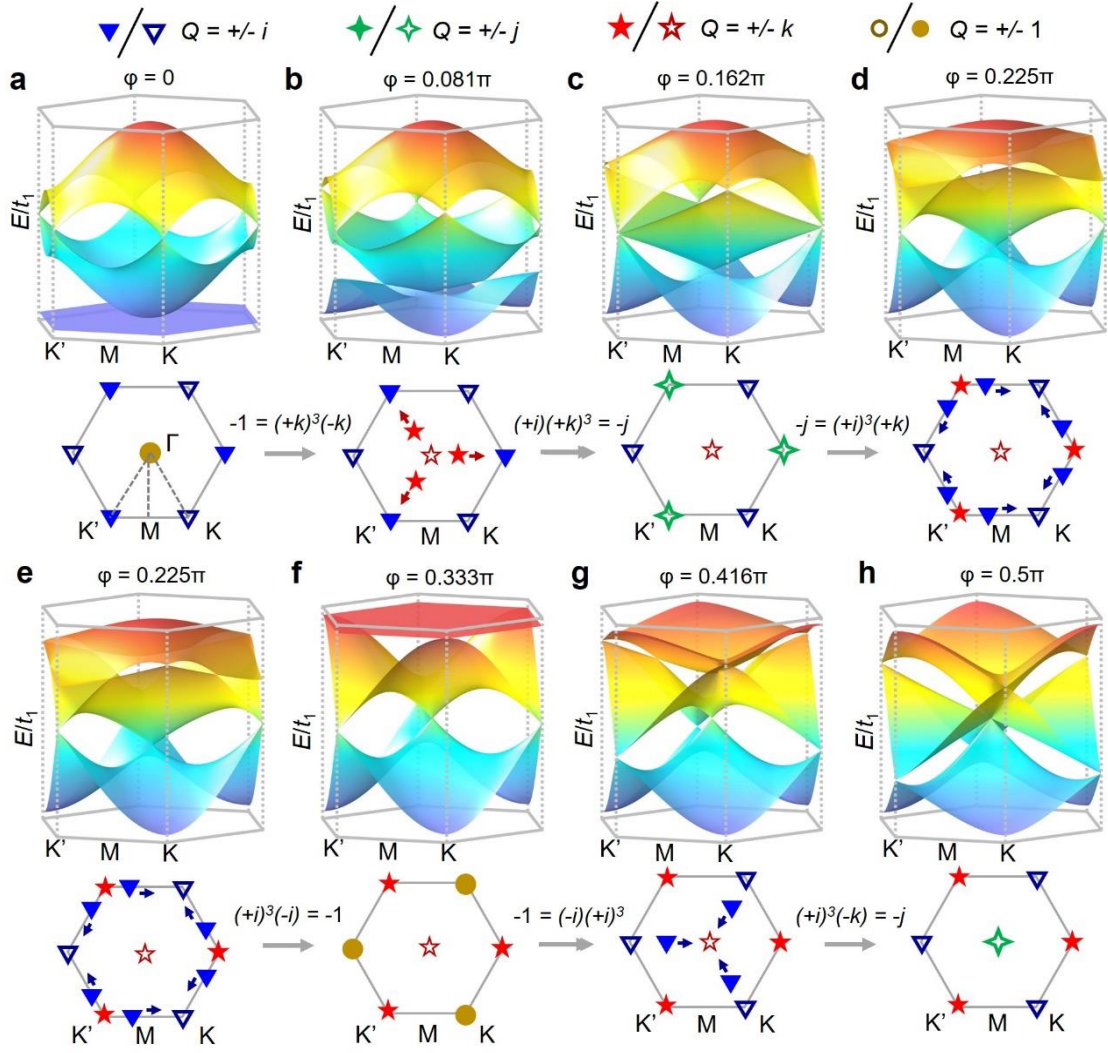

**Fig. S1. Evolution of the band structures (upper panels) and the non-Abelian nodes in gaps I and II (bottom panels) by continuously increasing magnetic flux strength. a-h**, Theoretical results from the diagram in Fig. 1c from $\varphi=0$ to $\varphi=0.5\pi$. Solid (open) blue triangles label the Dirac nodes in gap I with quaternion frame charge $\mathbb{Q}=i(-i)$. Solid (open) red quadrangular stars label the Dirac nodes in gap II with frame charge $\mathbb{Q}=j(-j)$. Solid (open) green pentagrams denote the linear triply-degenerate points with frame charges $\mathbb{Q}=k(-k)$. Solid (open) brown circles label the quadratic nodes with frame charge $\mathbb{Q}=-1(+1)$.

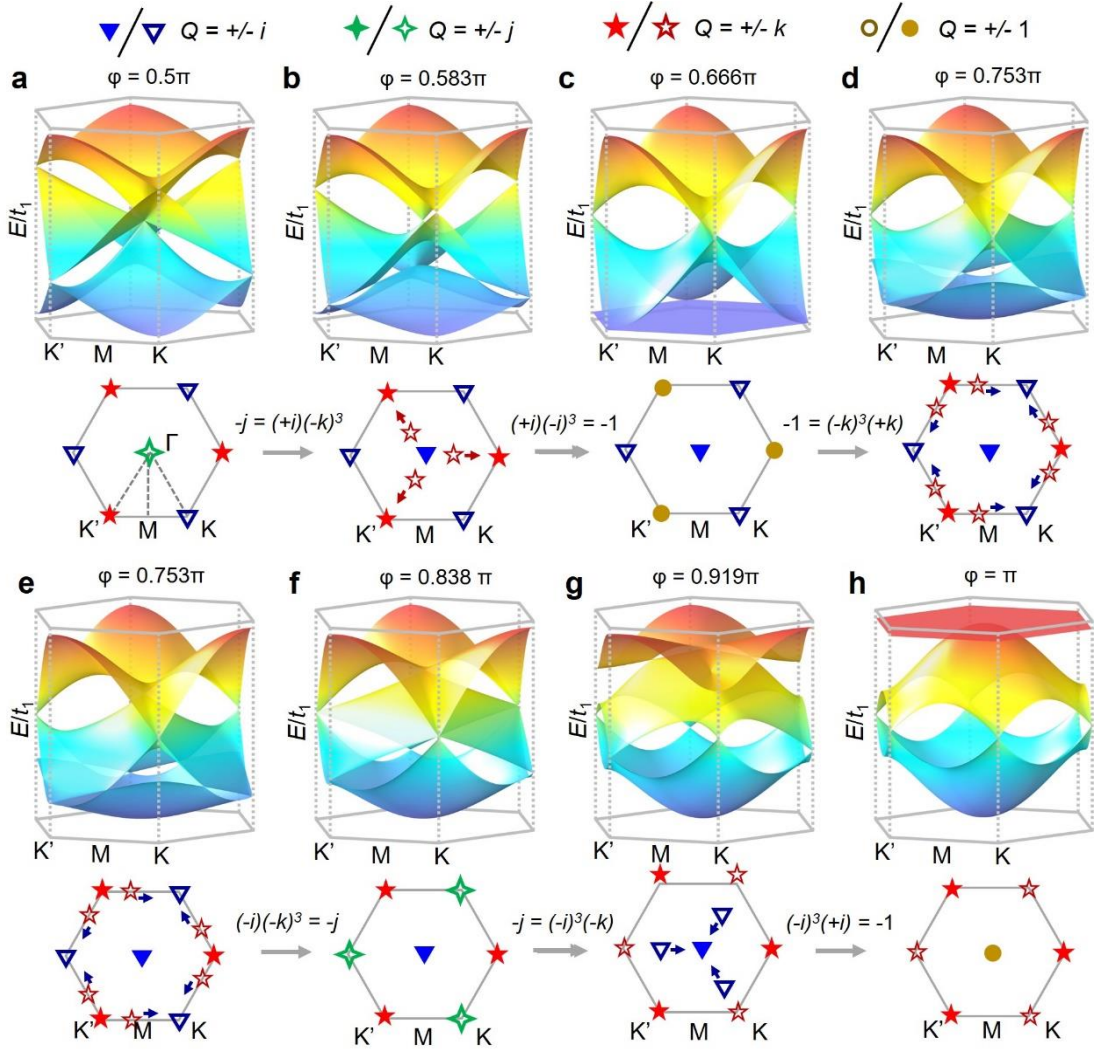

**Fig. S2. Evolution of the band structures (upper panels) and the non-Abelian nodes in gaps I and II (bottom panels) by continuously increasing magnetic flux strength. a-h,** Theoretical results from the diagram in Fig. 1c from $\varphi = 0.5\pi$ to $\varphi = \pi$. Solid (open) blue triangles label the Dirac nodes in gap I with quaternion frame charge $\mathbb{Q} = i(-i)$. Solid (open) red quadrangular stars label the Dirac nodes in gap II with frame charge $\mathbb{Q} = j(-j)$. Solid (open) green pentagrams denote the linear triply-degenerate points with frame charges $\mathbb{Q} = k(-k)$. Solid (open) brown circles label the quadratic nodes with frame charge $\mathbb{Q} = -1(+1)$.

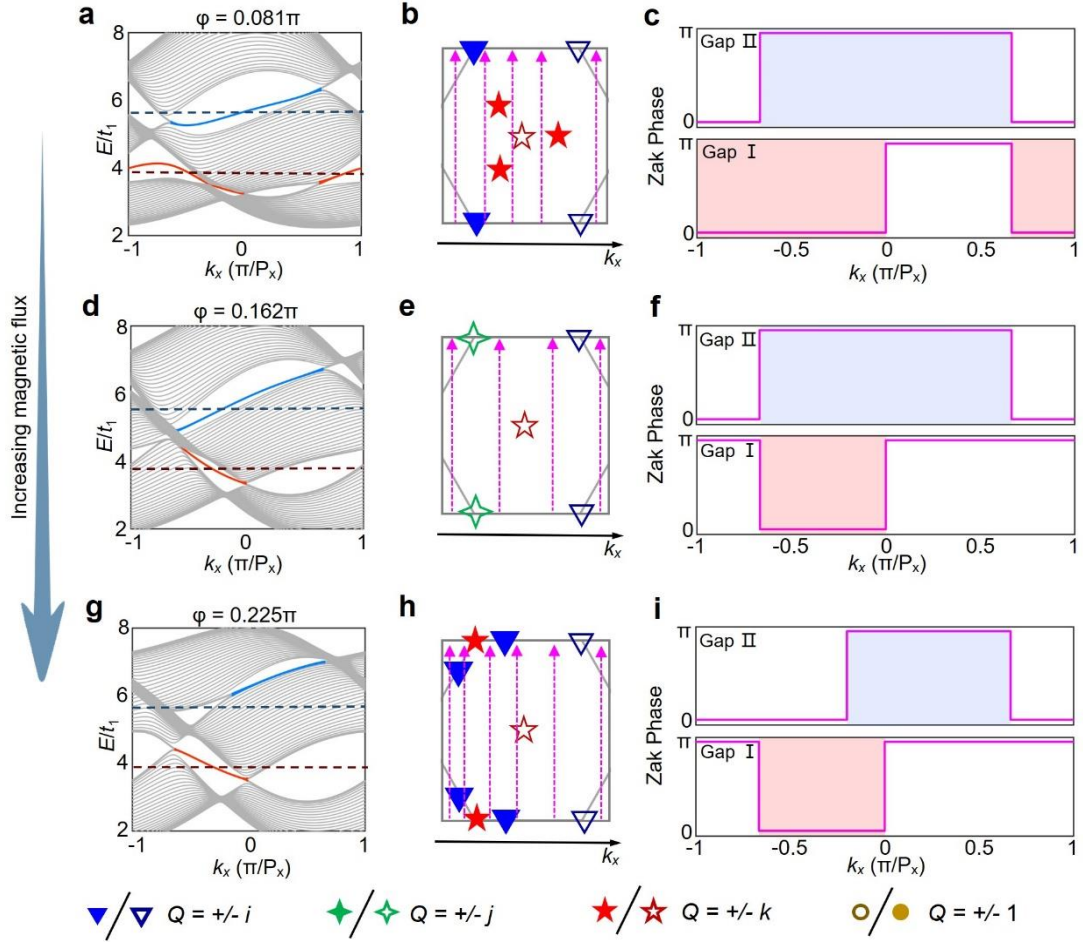

**Fig. S3. Relationship between multiband edge states and Zak phases of 1D subsystems. a**, **d**, **g**, Projected energy dispersions for structures that are finite in the x-direction but periodic in the y-direction. Gray curves represent bulk states blue (red) ones represent edge states in gap II (I). **b**, **e**, **h**, The topological node arrangement within Gaps II (I) throughout the entire Brillouin zone. Purple dashed arrows represent 1D $k_y$-loops sketched for calculating Zak phases. Zak phases may change when the loop bypass band nodes. **c**, **f**, **i**, Calculated Zak phase for 1D subsystems along $k_y$-loops by varying $k_x$. Red (blue) regions mark the appearance of edge state within the Gaps I (II).

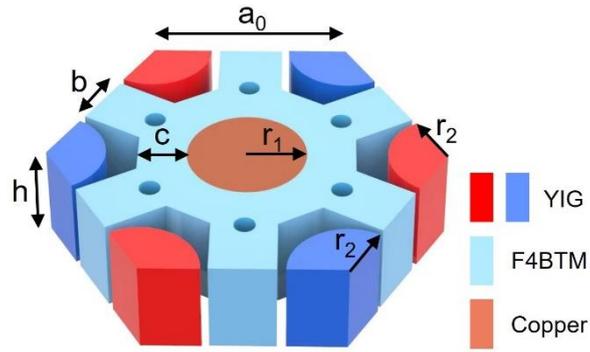

**Fig. S4. Unit cell of a time-reversal broken Kagome photonic crystal.** The structure parameters are b = 6 mm, c = 2.5 mm, $r_1$ = 5.4 mm, $r_2$ = 3.3 mm and the lattice constant $a_0$ = 17.5 mm. The thickness in the $z$-direction is h = 5 mm. The F4BTM dielectric material has a relative permittivity of 3.5. The radius of holes in dielectric material is 1 mm.